\documentclass[
 reprint,
 amsmath, amssymb,
 pra,
 aps,
 superscriptaddress
]{revtex4-2}

\usepackage{graphicx}
\usepackage{subfigure}
\usepackage{hyperref}
\usepackage{bm}
\usepackage{float}
\usepackage{color}
\usepackage{amsmath}
\usepackage{braket} 
\usepackage{ulem}

\DeclareMathOperator{\Tr}{Tr}

\newcommand{\inner}[2]{\left \langle #1 \middle| #2 \right \rangle}

\newcommand\va[1]{\boldsymbol{#1}}
\newcommand\expval[1]{\braket{#1}}
\newcommand\abs[1]{\vert{#1}\vert}

\hypersetup{colorlinks=true, citecolor=blue, urlcolor=blue, linkcolor=blue}
\begin{document}

\preprint{APS/123-QED}
\title{Switching Dynamics of Metastable Open Quantum Systems}

\author{Ya-Xin Xiang}
      \affiliation{National Laboratory of Solid State Microstructures and School of Physics,
	Collaborative Innovation Center of Advanced Microstructures, Nanjing University, Nanjing 210093, China}
\author{Weibin Li}
\affiliation{
School of Physics and Astronomy, and Centre for the Mathematics and Theoretical Physics of Quantum Non-equilibrium Systems, University of Nottingham, Nottingham, NG7 2RD, UK}
\author{Zhengyang Bai}
\email{zhybai@nju.edu.cn}
\affiliation{National Laboratory of Solid State Microstructures and School of Physics,
	Collaborative Innovation Center of Advanced Microstructures, Nanjing University, Nanjing 210093, China}

\author{Yu-Qiang Ma}
\email{myqiang@nju.edu.cn}
\affiliation{National Laboratory of Solid State Microstructures and School of Physics,
	Collaborative Innovation Center of Advanced Microstructures, Nanjing University, Nanjing 210093, China}
\affiliation{Hefei National Laboratory, Hefei 230088, China}
\begin{abstract}
Classical metastability manifests as noise-driven switching between disjoint basins of attraction and slowing down of relaxation, quantum systems like qubits and Rydberg atoms exhibit analogous behavior through collective quantum jumps and long-lived Liouvillian modes with a small spectral gap.
Though any metastable mode is expected to decay after a finite time, stochastic switching persists indefinitely. 
Here, we elaborate on the connection between switching dynamics and quantum metastability through the lens of the large deviation principles, spectral decomposition, and quantum-jump simulations. Specifically,  we distinguish the trajectory-level noise-induced metastability (stochastic switching) from the spectrum-level deterministic metastability (small Liouvillian gap) in a Markovian open quantum system with bistability. Without stochastic switching, whether a small spectral gap leads to slow relaxation depends on initial states. In contrast, with switching, the memory of initial conditions is quickly lost, and the relaxation is limited by the rare switching between the metastable states. 
Consistent with the exponential scaling of the Liouvillian gap with system size, the switching rates conform to the Arrhenius law, with the inverse system size serving as the nonequilibrium analog of temperature. Using the dynamical path integral and the instanton approach, we further extend the connection between the quasipotential functional and the probabilities of rare fluctuations to the quantum realm. These results provide new insights into quantum bistability and the relaxation processes of strongly interacting, dissipative quantum systems far away from the thermodynamic limit.
\end{abstract}

\maketitle

\section{introduction}

Spontaneous switching between distinct metastable states in driven-dissipative quantum many-body systems (known as collective quantum jumps~\cite{risken1988quantum,lapidus1999stochastic,lee2012collective}) is recently receiving
growing attention in quantum science as a signature of nonequilibrium phase transitions~\cite{fink2017observation, fink2017signatures,ricci2017optically, rondin2017direct, foss2017emergent,mavrogordatos2018rare,andersen2020quantum,stochastic2020he,muppalla2018bistability}, with potential applications in quantum information~\cite{marvrogordatos2017spontaneous,minganti2016exact} and metrology~\cite{wu2024nonlinearity,li2024collective}.
At first glance, the existence of such collective jumps in Markovian open quantum systems presents a paradox. 
These systems are governed by the Lindblad equation~\cite{gorini1976complete,lindblad1976} and are expected to relax toward the unique stationary state as time approaches infinity~\cite{nigro2019uniqueness}. 

Despite the small Liouvillian spectral gap and the onset of metastability associated with first-order dissipative phase transitions ~\cite{minganti2018spectral,macieszczak2021theory,macieszczak2016towards,vogel1988quatum,casteels2016power,casteels2017critical,vicentini2018,chen2023quantum,li2024emergent, carr2013Nonequilibrium,ding2020phase}, metastable modes in Liouville space inevitably decay over time in finite systems.  This behavior is in stark contrast to persistent collective jumps observed in bistable systems~\cite{lee2012collective,fink2017observation, fink2017signatures}.  

This apparent contradiction not only highlights the fundamental distinction between the trajectory-level metastability (stochastic switching) and the spectrum-level metastability (small Liouvillian gap), but also raises a crucial question regarding the relation between mean-field (MF) descriptions of phase transitions and quantum metastability in finite systems. 
In finite systems, quantum fluctuations transform stable MF fixed points into metastable states with finite lifetimes.
These distinct metastable states are mixed through collective quantum jumps, in much the same way as the coexisting classical basins of attraction are connected by large deviations (LDs), contrasting with the small, typical fluctuations within a single attractive basin~\cite{Dykman1994LargeFA,zakine2023minimum,touchette2009large}.

\begin{figure}[t]
     \centering
	\includegraphics[width = 8.3 cm, keepaspectratio]{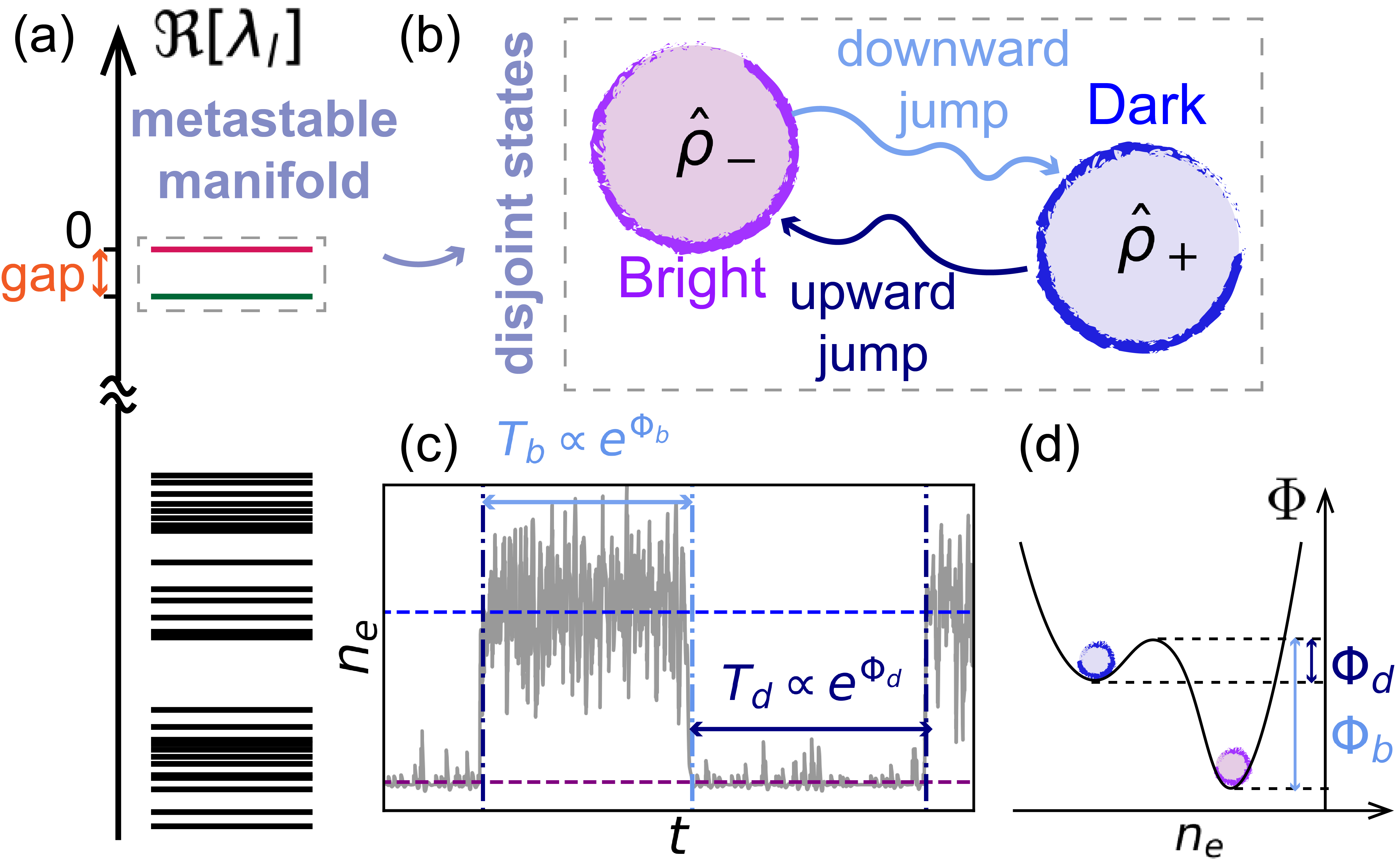}
	\caption{Sketch of quantum metastability and collective quantum jump. 
    (a) Quantum metastability arises from both a small spectral gap between a low-lying eigenmode (green line) and the unique stationary state (red line) and a separation between them and the remainder (black lines) in the real part of the Liouvillian spectrum. The first two low-lying modes belong to the (metastable manifold) MM, which is spanned by
    (b) two disjoint metastable states $\hat{\rho}_+$ and $\hat{\rho}_-$  corresponding to the two stable MF fixed points in the thermodynamic limit. The two metastable states are mixed through
    (c) upward and downward collective jumps with rates $T_d^{-1},T_b^{-1}$ that follow an Arrhenius-type scaling $T_{b,d}^{-1}\propto e^{-\Phi_{b,d}}$, where
    (d) $\Phi_{b},\Phi_{d}$ are the effective energy barriers quantifying the relative stability between the bright ($b$) and dark ($d$) states. The purple and blue horizontal lines indicate the Rydberg population $n_e$ of the two stable MF fixed points, while dark and light blue vertical lines in (c) mark the upward and downward collective jumps, respectively.}
	\label{fig:Scheme}
\end{figure}

However, a conceptual gap persists concerning the connection between the two facets of quantum metastability: how the metastable states correspond to the long-lived states spanning the metastable manifold (MM) of Liouville space~\cite{letsher2017bistability,brookes2021critical,fitzpatrick2017observation,macieszczak2021theory}, and how the single spectral gap governs the transition rates between these metastable states.
Classical bistable systems typically exhibit a preference for one metastable state over the other, with the lifetimes of these states following the Arrhenius law~\cite{landau1980stat,bertini2015mft,Dykman1994LargeFA}. Both the Gibbs-Boltzmann distribution and the theory of first-passage times predict that mean switching times between these metastable states increase exponentially with the free energy (or its nonequilibrium counterpart, the quasipotential)  barriers~\cite{hanggi1990reaction,kampen2007stochastic,gardiner2009stochastic}. 
It is natural to wonder whether this rule applies to bistable systems subjected to quantum fluctuations at zero temperature. 
If this is the case, the lifetimes of the two quantum metastable states should exhibit an Arrhenius-type exponential scaling with the effective energy barrier, mirroring their equilibrium counterparts. 

In this work, we demonstrate these insights by exploring the switching dynamics of a many-body open quantum system with bistability~\cite{lee2011Antiferromagnetic,carr2013Nonequilibrium}.
We will focus our study with the Rydberg atom setting, which has been widely adopted as a paradigmatic platform for exploring complex many-body dynamics due to its strong and controllable atomic interactions~\cite{browaeys2020many,sibalic2018rydberg,weimer2010rydberg}.
MF analysis predicts a first-order dissipative phase transition between states with high and low Rydberg populations, referred to as the bright and dark states, respectively. The bright state exhibits a high photon emission rate, while the dark state yields a suppressed photon emission rate~\cite{gardiner1992wave,plenio1998the,carr2013Nonequilibrium,liu2024Emergence}.   

In finite systems far away from the thermodynamic limit, discontinuous transitions manifest as metastability, producing distinctive signatures in the statistics of quantum fluctuations.  The LD function of the photon emission rates, derived from the spectral analysis of tilted Liouvillian operators, reveals singularities originating from metastable states with different Rydberg populations~\cite{juan2010thermodynamics}. 
Additionally, the system is characterized by a real Liouvillian gap that decreases exponentially with increasing system size (number of atoms) [see Fig.~\ref{fig:Scheme}(a)]. 
This allows us to extract the two disjoint metastable states that span the MM through spectral decomposition, leveraging the trace-preserving symmetry of the Lindbladian. The resulting disjoint metastable states $\hat{\rho}_-$ and $\hat\rho_+$ in Liouville space are found to directly correspond to their MF counterparts and are consequently identified as the quantum bright and dark states [see Fig.~\ref{fig:Scheme}(b)]. 
Driven by quantum fluctuations, the system switches back and forth between the dark and bright states on timescales significantly exceeding those of microscopic dynamics [see Fig.~\ref{fig:Scheme}(c)]. Through the lens of LDs, the lifetimes of metastable states show an exponential dependence on the effective energy barriers [Fig.~\ref{fig:Scheme}(d)], thereby linking the probabilities of large fluctuations to the steady-state occupation probabilities of the two states~\cite{zakine2023minimum,elgart2004rare}.

In the absence of free energy functional for nonequilibrium states, we resort to the dynamical path integral~\cite{Martin1973,Janssen1976,Tauber2014critical} combined with the instanton approach~\cite{zakine2023minimum,elgart2004rare} to determine the effective energy barriers. This allows us to identify the inverse of the system size as the nonequilibrium analog of temperature, which also serves as the LD parameter. The suppressed switching in enlarged systems mirrors the findings of previous studies~\cite{carr2013Nonequilibrium,ding2020phase,biondi2017nonequilibrium,huybrechts2020dynamical,rodriguez2017probing,sett2024emergent,vukics2019finite,casteels2017critical}.
To further pinpoint the impact of switching dynamics on the slowest relaxation time, we directly measure the waiting times of collective jumps from the trajectories through the quantum-jump simulations for various system sizes and control parameters. In line with the predictions from the instanton approach, the waiting times show the same exponential size scaling as the Liouvillian gap. 

The suppression of relaxation by quantum bistability results from metastability at both the trajectory and spectrum levels. These two facets are intertwined: stochastic switching between the two states inevitably excites the low-lying mode associated with the small spectral gap.  In contrast, in a far-off resonant regime, this system manifests metastability without bistability, where the stationary state is entirely confined to one of the two disjoint subspaces.  In the absence of bistability and switching, whether the small spectral gap leads to a controllable relaxation time depends on the initial condition~\cite{zhang2025observation,carollo2021exp}. By extending the classical rare-event statistics to the quantum realm, our findings offer new insights into quantum bistability while systematically revealing the relaxation dynamics in strongly interacting, dissipative quantum systems.

The paper is organized as follows. In Sec.~\ref{sec:DPT}, we first introduce the physical model (Sec.~\ref{sec:model}), derive the MF equations of motion, and obtain the MF phase diagram (Sec.~\ref{sec:MF}). We then analyze the relation between MF fixed points and quantum metastable states in finite systems by examining both the statistics of quantum fluctuations (Sec.~\ref{sec:LD}) and the low-lying eigenmatrices of the Lindblad superoperator (Sec.~\ref{sec:ED}). After establishing the connection between long-time dynamics and switching between quantum metastable  states, we investigate their steady-state occupation probabilities and discuss the impacts of switching (and the lack thereof) on the slowest relaxation (Sec.~\ref{sec:PDF}). 
In Sec.~\ref{sec:switching}, we begin by estimating the effective energy barriers and the switching times using the dynamical path integral and the instanton approach (Sec.~\ref{sec:instanton}). We then employ the quantum-jump Monte-Carlo simulations to extract the mean waiting times and the effective energy barriers (Sec.~\ref{sec:QJMC}). We end this section with a comparison of effective energy barriers and the estimated relaxation times across all the aforementioned methods (i.e., the spectral method, the instanton approach, and the quantum-jump simulations). 
The paper concludes with summary and discussion in Sec.~\ref{sec:conclusion}.

\section{\label{sec:DPT}Dissipative discontinuous phase transition}

\subsection{\label{sec:model}The physical system}

In our setting, we consider an ensemble of $N$ atoms consisting of two electronic states. Each atom (indexed by subscript $l$) is continuously laser-excited  from the ground state $\ket{g}_l$ to electronically excited Rydberg state $\ket{e}_l$ with Rabi frequency $\Omega$ and detuning $\Delta$ from resonance. Once excited, Rydberg states interact via strong and long-range interactions. Microscopic dynamics of the system is described by the Lindblad equation ($\hbar=1$) 
\begin{equation}
\label{eq:lindblad}
\hat{\mathcal{L}}[\hat{\rho}]\equiv\partial_t\hat{\rho}=-i[\hat{H},\hat{\rho}]+\hat{\boldsymbol{D}}[\hat{\rho}]
\end{equation}
Within the rotating-wave approximation, the resultant Hamiltonian $\hat{H}$ in the interaction picture reads
\begin{equation}
\label{eq:H}
\hat{H} = \sum_l{\left[-\Delta\hat n_l+\Omega\hat{\sigma}^{x}_l+\frac{V}{N-1}\sum_{l < k}\hat n_k\hat n_l\right]}
\end{equation}
where $\hat{\sigma}_l^{\alpha}~(\alpha=x, y, z)$ denote the Pauli matrices acting on the $l$-th atom, and $\hat n_{l} =\hat{\sigma}_l^{z}+\mathbf{I}_l/2$ represents the Rydberg number operator.

The last term of the Hamiltonian~(\ref{eq:H}) represents the Rydberg interactions between the $l$-th and $k$-th atoms.

Depending on the selected Rydberg states, excited atoms separated by distance $R$ can participate in dipole-dipole ($\sim R^{-3}$) and van de Waals ($\sim R^{-6}$) interactions \cite{sibalic2018rydberg,browaeys2020many,saffman2010quantum}. To investigate the long-time dynamics of large systems, we approximate the interactions $V$ as a constant all-to-all coupling, normalized to render the Hamiltonian extensive.
This approximation is valid for high-dimensional Rydberg gases~\cite{lee2012collective, carr2013Nonequilibrium}.

Each atom couples with the local environment, which is always in the vacuum state, resulting in a finite lifetime and the dissipative dynamics
\begin{equation}
\label{eq:diss}
\hat{\boldsymbol{D}}\left[\hat{\rho}\right]=\frac{\gamma}{2}\sum_{l}{\left(2\hat{\sigma}_l^{-}\hat{\rho}\hat{\sigma}_l^{+}-\left\{\hat{\sigma}_l^{+}\hat{\sigma}_l^{-},\hat{\rho}\right\}\right)}
\end{equation}
where the operators 
$\hat{\sigma}_l^\pm\equiv\hat{\sigma}_l^{x}\pm i\hat{\sigma}_l^{y}$ flips the atomic state.
Eq.~(\ref{eq:diss}) describes the decay processes from $\ket{e}$ to $\ket{g}$ with rate $\gamma$.

\subsection{\label{sec:MF}Mean-field model}

We begin with a MF ansatz $\hat{\rho}=\prod_{l}{\otimes\hat{\rho}_l}$  to characterize the dynamical behavior of this system in the thermodynamic limit. Introducing the collective spin operators ($\alpha=x, y, z$) $\hat{S}^\alpha=\sum_l{\hat{\sigma}^\alpha_l}$, we obtain the following set of coupled nonlinear equations for the expectation values $m_\alpha\equiv\Tr{[\hat{S}^\alpha\hat{\rho}]/S}$ (with the total spin $S=N/2$)
\begin{subequations}
\label{eq:EOM_MF}
\begin{align}
&\partial_t m_x = -\left[\frac{V}{2}( m_z+1)-\Delta\right]m_y - \frac{\gamma}{2}m_x \label{eq:dmxdt}\\
&\partial_t m_y = -\Omega m_z + \left[\frac{V}{2}(m_z+1)-\Delta\right]m_x - \frac{\gamma}{2}m_y\\
&\partial_t m_z = \Omega m_y - \gamma(m_z+1)  \label{eq:dmzdt}
\end{align}
\end{subequations}
where the Rydberg population $n_e\equiv\Tr[{\sum_l{\hat{n}_{l}}\hat{\rho}}]/N=(m^{z}+1)/2$.  Without loss of generality, we choose $\gamma^{-1}$ as the time unit and fix Rabi frequency $\Omega=1.5$ and atomic interaction $V=10$. 

\begin{figure}[b]
	\centering
	\includegraphics[width = 8.3 cm, keepaspectratio]{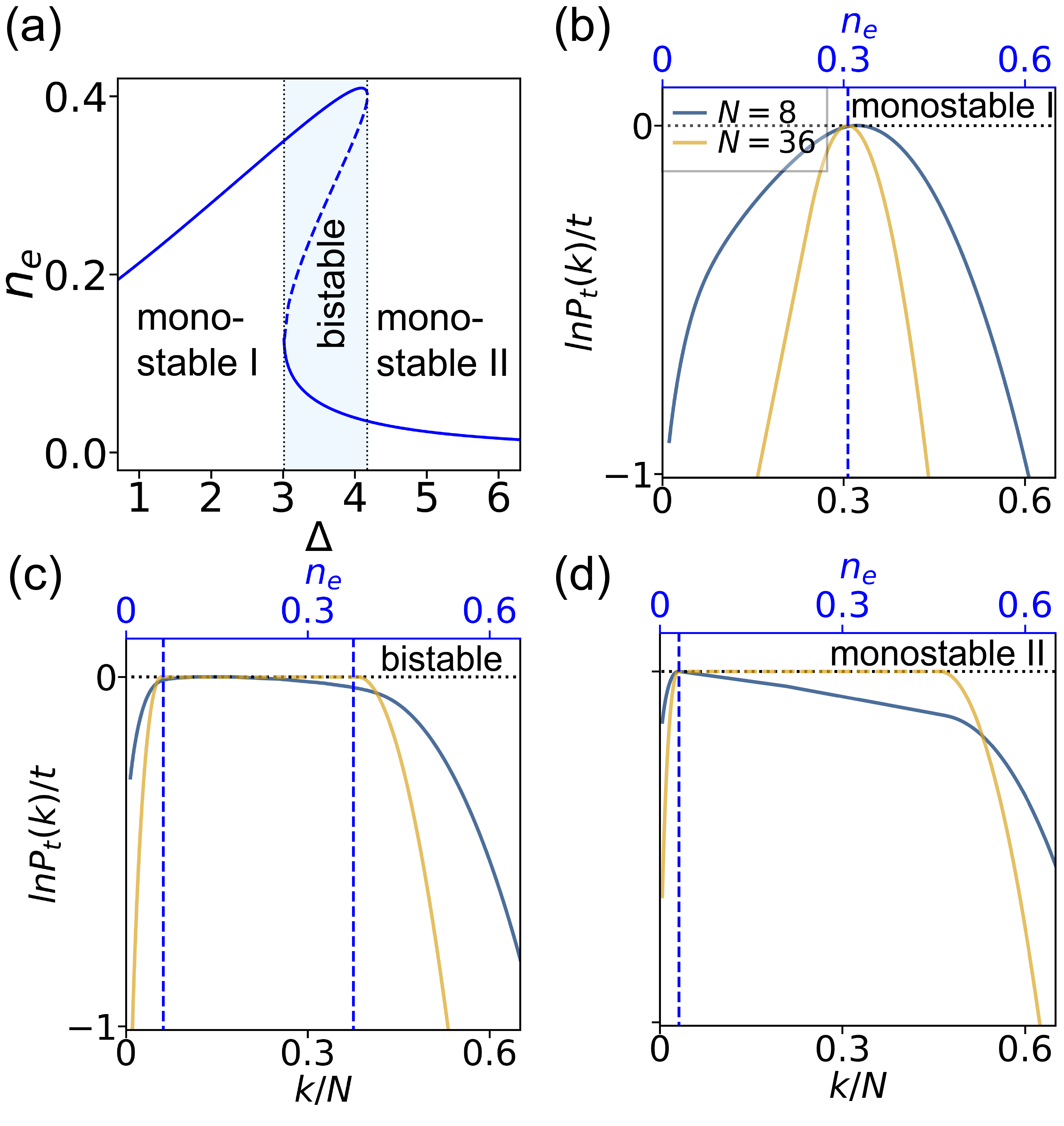}
	\caption{(a) MF stable (unstable) fixed points as a function of detuning $\Delta$ represented by solid (dashed) lines. Phase boundaries are highlighted by dotted gray lines, and the bistable region in between is shaded blue. 
	Statistics of time-averaged photon emission rate in systems with different number of atoms (color-coded) obtained within the quantum LD formalism for (b) $\Delta=2.4$, (c) $\Delta=3.4$, and (d) $\Delta=4.4$. The rise of two maxima connected by vertical dashed lines indicates kink of the LD functions. The blue dashed lines represent the excitation population of the MF stable fixed points.}
	\label{fig:DPT_MF_LD}
\end{figure}

Linear stability analysis of the fixed points derived from Eqs~(\ref{eq:dmxdt})-(\ref{eq:dmzdt}) reveals that the system exhibits a discontinuous phase transition as a function of detuning. As illustrated in Fig. \ref{fig:DPT_MF_LD}(a), based on the number of stable fixed points, the phase diagram is partitioned into three distinct regimes: monostable I, monostable II, and bistable. In the bistable regime, each of the two stable fixed points possesses a unique basin of attraction, and the system evolves towards either of the two stable fixed points, determined by the initial conditions.

Phase transitions strictly occur in the thermodynamic limit, in finite systems, they are instead characterized by the emergence of metastability. Quantum fluctuations destabilize the MF fixed points, replacing them with metastable quantum states of finite lifetimes.
As demonstrated in Fig. ~\ref{fig:Scheme}(c), collective quantum jumps between two metastable states arise in the parameter regime where the MF model displays bistability ($\Delta\approx3\sim4.2$) ~\cite{lee2012collective}. 

\subsection{\label{sec:LD}Quantum metastability in finite systems}

Signatures of discontinuous dissipative phase transitions in finite systems also manifest in the quantum-jump statistics, as trajectories in different metastable states exhibit different collapse and phonton emission rates~\cite{plenio1998the,gardiner1992wave,dalibard1992wave}. 
In the long-time limit $t\gg 1$, the probability $P_t(K)$ of observing $K$ emitted photons adopts a LD form~\cite{juan2010thermodynamics} 
\begin{equation}
P_t(K)\approx e^{-t\phi(k)}
\end{equation}
with $k=K/t$ being the time-averaged photon emission rate. 

The LD function $\phi(k)$ contains all information about the probability of $K$ at long times. 
The statistics of $K$ is described via the generating function,
which also acquires a LD form, i.e., 
\begin{equation}
Z_t(s)\equiv\sum_{K=0}^{\infty}{P_t(K)e^{-sK}}\approx e^{t\theta(s)}
\end{equation}
Here $s$ is the conjugate field to the dynamic observable $K$, and the two LD functions $\theta(s),\phi(k)$ are related by a Legendre transform~\cite{touchette2009large,juan2010thermodynamics}, $\theta(s)=-\min_k{[\phi(k)+ks]}$. 

We obtain the LD function $\theta(s)$ by finding directly the largest eigenvalue of the tilted generator~\cite{touchette2009large,causer2020dynamics,whitelam2021varied,ates2012dynamical},
\begin{equation}
\hat{\mathcal{L}}_s[\hat{\rho}]=i[\hat{\rho},\hat{H}_\text{eff}] + e^{-s}\hat{L}\hat{\rho}\hat{L}^\dagger
\end{equation}
Since the all-to-all coupling preserves the permutation symmetry, we rewrite Hamiltonian~\eqref{eq:H} and dissipation \eqref{eq:diss} in terms of collective Dicke states $\ket{M}$~\cite{dicke1954coherence}. 
Thus, we obtain the effective non-Hermitian Hamiltonian $\hat{H}_\text{eff}=\Omega\hat{S}^x-\frac{V}{2N}\hat{S}^{+}\hat{S}^{-}+(\frac{V-i\gamma}{2}-\Delta)\hat{S}^z-iS/2$ and the jump operator $\hat{L}=\sum_{M}{\sqrt{M+S}\ket{M-1}\bra{M}}$ after dropping irrelevant constant terms and focusing on large $N$, where the operators $\hat{S}^{\pm}=\sum_{l}{\hat{\sigma}_l^{\pm}}$ satisfy $\hat{S}^{\pm}\ket{M}=\sqrt{\left(S\mp M\right)\left(S\pm M+1\right)}\ket{M\pm 1}$ and the commutation relation $[\hat{S}^+,\hat{S}^-]=2\hat{S}^z$.

Because photon emission is attributed to the spontaneous decay of excited atoms, at late times $t\gg 1$, the average number of emitted photons approaches the steady-state excitation population, i.e. $\expval{K}/t \approx N n_e$. As shown in Figs.~\ref{fig:DPT_MF_LD}(b)-(d), in monostable I regime [$\Delta=2.4$; see panel (b)], the LD function $-\phi(k)\equiv\lim_{t\to+\infty}{\ln{P_t}(K)/t}$ peaks and vanishes at the typical value that corresponds to the MF Rydberg densities. As the number of atoms $N$ increases, the distributions become narrower, consistent with the law of large numbers. In the presence of bistability [$\Delta=3.4$; see panel (c)], the LD function becomes bimodal, vanishing completely between two maxima, corresponding to high and low excitation populations.
This originates from singularities (kinks) of the LD function $\theta(s)$ and is indicative of distinct dynamical phases~\cite{causer2020dynamics,carllo2020entanglement,juan2010thermodynamics,whitelam2021varied,garrahan2007dynamical}. 
Hereafter, we designate the metastable state with higher (lower) excitation populations as the bright (dark) state, reflecting the higher photon emission rate of the bright state relative to the dark state.

\begin{figure}[b]
	\centering
	\includegraphics[width = 8.3 cm, keepaspectratio]{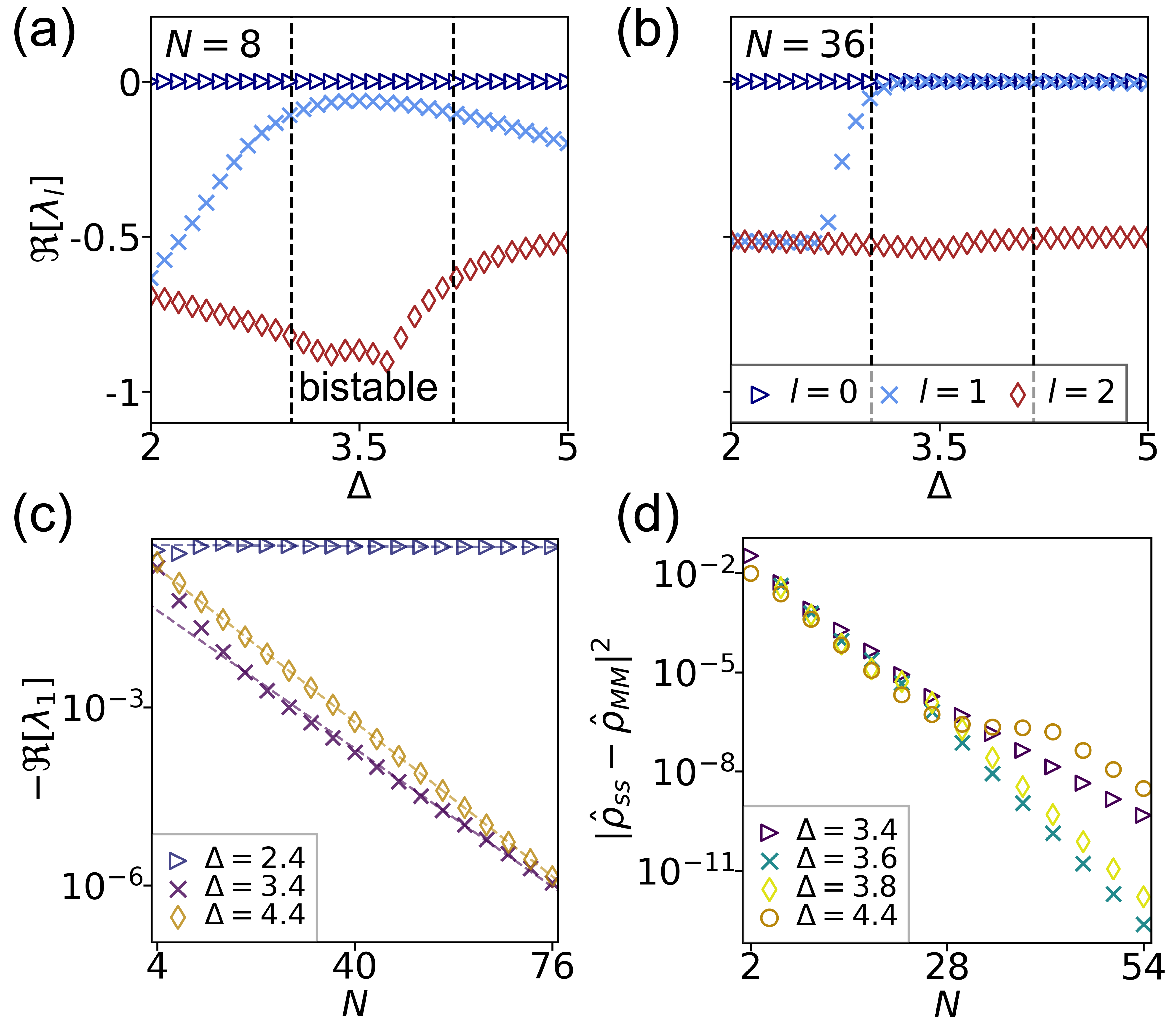}
	\caption{Real part of the Liouvillian eigenvalues for (a) $N=8$ and (b) $N=36$. The index $l$ labels the eigenvalues. The Liouvillian eigenvalues $\lambda_l$ are arranged by their real parts in descending order, and $\lambda_0=0$. The bistable regime is located between the two black dashed lines.
	(c) Finite-size scaling for the spectral gap in monostable I ($\Delta$=2.4), bistable ($\Delta=3.4$), and monostable II ($\Delta=4.4$) regimes. Dashed lines are obtained from fitting $-\Re[\lambda_1]=b e^{a N}$. 
	(d) The errors (where $\abs{\hat{A}}^2\equiv \expval{\hat{A}^\dagger,\hat{A}}$ and $\expval{\hat{A},\hat{B}}\equiv\Tr{[\hat{A}\hat{B}]}$) of approximation of the stationary state $\hat{\rho}_\text{ss}$ to the linear combination of the metastable disjoint subspaces $\hat{\rho}_\text{MM}\equiv \sum_{\alpha=\pm}{D[\hat{\rho}_\alpha,\hat{\rho}_\text{ss}]\hat{\rho}_\alpha}$ with coefficients being the normalized Hilbert-Schmidt inner products $D[\hat{A},\hat{B}]\equiv\expval{\hat{A}^\dagger,\hat{B}}/\expval{\hat{A}^\dagger,\hat{A}}$ in the bistable ($\Delta=3.4,3.6,3.8$) and monostable II ($\Delta=4.4$) regimes.}
	\label{fig:DPT_SP}
\end{figure}

However, the kinks in the LD functions and the concomitant bimodal statistics persist even after further increasing the detuning, which is at odds with the monostability predicted by MF theory [$\Delta=4.4$; see panel (d)]. The discrepancy is attributed to the emergence of a diverging relaxation time in the absence of bistability, which also induces singularities in the LD functions~\cite{whitelam2021varied}. Therefore, spectral analysis suggests the onset of metastability at the spectrum level (small Liouvillian gap), which might not be associated with metastability at the trajectory level (bistability).

\subsection{\label{sec:ED}Quantum metastable states}

We explore the relation between the quantum metastability and bistability, through examining the low-lying modes of the Liouvillian superoperator. We obtain its eigenvalues and eigenmatrices $\hat{\mathcal{L}}$ via $\hat{\mathcal{L}}[\hat{\rho}_l]=\lambda_l\hat{\rho}_l$ and sort them according to $0=\lambda_0\geq\Re[\lambda_1]\geq...\geq\Re[\lambda_{(N+1)^2}]$. It follows from $\lambda_0=0$ that the stationary density matrix $\hat{\rho}_\text{ss}=\hat{\rho}_0/\Tr{[\hat{\rho}_0]}$, and the Liouvillian spectral gap is defined as $-\Re[\lambda_1]$. 

The real parts of the first three eigenvalues for small and large systems are displayed in Figs.~\ref{fig:DPT_SP}(a) and (b), respectively. For small systems (e.g., $N=8$), the spectrum is gapped for all values of $\Delta$. In contrast, for large systems (e.g., $N=36$), the spectral gap decreases with the detuning and almost vanishes after reaching the left boundary of the bistable regime. After plotting the gap in Fig.~\ref{fig:DPT_SP}(c), we find that the gap is independent of system size for small $\Delta$ (monostable I). However, for large $\Delta$ (bistable and monostable II), the gap decreases exponentially with $N$. This ultimately results in the emergent singularity in the LD functions shown in Figs.~\ref{fig:DPT_MF_LD}(c) and (d).

\begin{figure}[b]
	\centering
	\includegraphics[width = 8.3 cm, keepaspectratio]{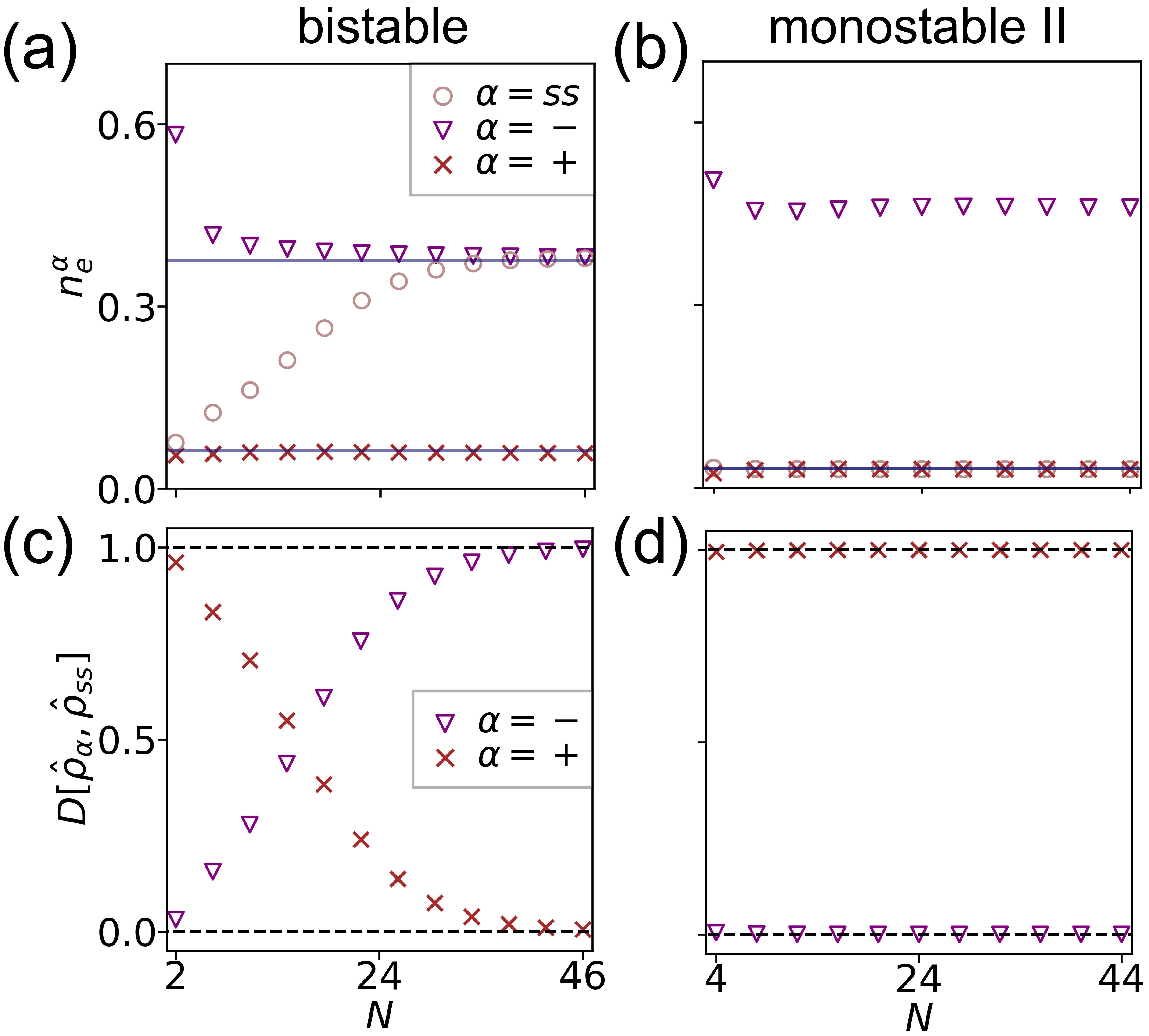}
	\caption{Average excited population according to the density matrices $\hat{\rho}_{\text{ss}}, \hat{\rho}_{\pm}$ as a function of $N$ for (a) $\Delta=3.4$ (bistable) and (b) $\Delta=4.4$ (monostable II). The blue lines represent the MF stable fixed points.
    The normalized Hilbert-Schmidt inner products $D[\hat{\rho}_{\pm},\hat{\rho}_{\text{ss}}]$ between $\hat{\rho}_{\pm}$ and the steady-state $\hat{\rho}_{\text{ss}}$ as a function of $N$ for (c) $\Delta=3.4$ and (d) $\Delta=4.4$. The black dashed lines represent the upper and lower bounds.}
    \label{fig:D3.4_vs_D4.4}
\end{figure}

The eigenvalue $\lambda_1$ is complex for small detuning and becomes real as it approaches the bistable regime [see Sec. I of supplementary material (SM)]. In this case, the separation in the lifetimes of the first two eigenmodes $\hat{\rho}_{\text{ss}}$ and $\hat{\rho}_{1}$ suggests a double degeneracy, whereby the MM is spanned by a pair of orthonormal basis (density matrices), denoted as $\hat{\rho}_\pm$ (with $\Tr{[\hat{\rho}_\pm]}=1$ and $\hat{\rho}_\pm=\hat{\rho}_\pm^\dagger$). It follows that $\hat{\rho}_1=a_+\hat{\rho}_{+}+a_{-}\hat{\rho}_{-}$. Because the Lindblad equation preserves the trace (probabilities), any eigenmatrices with nonzero eigenvalues are traceless. Thus, we obtain $a_{+}=-a_{-}$.
In addition, it follows from the Lindblad equation~\eqref{eq:lindblad} that for any Liouvillian, if $\hat{L}[\hat{\rho}] = \lambda\hat{\rho}$, then $\hat{L}[\hat{\rho}^\dagger] = \lambda^\ast\hat{\rho}^\dagger$. Therefore, when $\lambda_1$ is real and of degeneracy $1$, the eigenmatrix $\hat{\rho}_1$ is Hermitian (i.e., $a_{\pm}\in\mathbb{R}$) and can be diagonalized, yielding the spectral decomposition~\cite{minganti2018spectral}
\begin{equation}
\hat{\rho}_1 =\sum_{l}{\theta(\alpha_l)\alpha_l\ket{\alpha_l}\bra{\alpha_l}} -\sum_{l}{\theta(-\alpha_l)(-\alpha_l)\ket{\alpha_l}\bra{\alpha_l}} 
\end{equation}
where $\theta(x)$ is the Heaviside step function, and all eigenvalues $\alpha_l$ are real and $\inner{\alpha_l}{\alpha_k} = \delta_{l,k}$. 

One immediately recognizes the two density matrices $\hat{\rho}_{\pm}$ as the subspaces spanned by the eigenvectors (pure states) of $\hat{\rho}_1$ with positive ($+$) and negative ($-$) eigenvalues, i.e.,
\begin{subequations}
\label{eq:SD}
\begin{align}
\hat{\rho}_- & \propto \sum_{l}{\theta(-\alpha_l)(-\alpha_l)\ket{\alpha_l}\bra{\alpha_l}} \\
\hat{\rho}_+ & \propto\sum_{l}{\theta(\alpha_l)\alpha_l\ket{\alpha_l}\bra{\alpha_l}}
\end{align}
\end{subequations}
followed by normalization ensuring $\Tr{[\hat{\rho}_\pm]}=1$.  
We recognize that the two subspaces are by construction orthogonal, i.e., $\expval{\hat{\rho}_+,\hat{\rho}_-} =0$ (where the Hilbert-Schmidt inner product (for operators $\hat{A}$ and $\hat{B}$)  $\expval{\hat{A},\hat{B}}\equiv\Tr{[\hat{A}\hat{B}]}$), and therefore are also the disjoint subspaces of the MM associated with the slowest relaxation~\cite{minganti2018spectral,macieszczak2016towards,macieszczak2021theory}.
We project the stationary state $\hat{\rho}_{ss}$ onto the MM through $\mathcal{P}_\text{MM}[\hat{\rho}_\text{ss}]\equiv\hat{\rho}_\text{MM}=\sum_{\alpha=\pm}{D[\hat{\rho}_\alpha,\hat{\rho}_\text{ss}] \hat{\rho}_\alpha}$, where the normalized inner product $D[\hat{A},\hat{B}]\equiv\expval{\hat{A}^\dagger, \hat{B}}/\expval{\hat{A}^\dagger, \hat{A}}$. As sketched in Fig.~\ref{fig:DPT_SP}(d), in both the bistable and monostable II regimes, the error $\abs{\hat{\rho}_\text{ss}-\hat{\rho}_\text{MM}}^2$ is small and also decreases exponentially
with $N$. 

The relation between the two metastable subspaces $\hat\rho_\pm$ and the two stable MF fixed points is elucidated by examining the average excitation densities ($\alpha=\text{ss},\pm$) $n_e^{\alpha}=\Tr{[\hat{n}_e\hat{\rho}_{\alpha}]}$. Within the bistable regime [$\Delta=3.4$;  Figs.~\ref{fig:D3.4_vs_D4.4}(a) and (c)], the two metastable states $\hat\rho_\pm$ are characterized by distinct excitation densities that correspond to those of the dark and bright states. Therefore, we designate the states $\hat{\rho}_{+}$ and $\hat\rho_{-}$ as \textit{the quantum dark and bright states}, respectively.  
The steady-state excitation population is a mixture of the two metastable states and asymptotically approaches the bright (dark) state for small (large) detuning when $N\to\infty$. This suggests that, in large systems, the stationary state emerges as a statistical mixture of the quantum dark and bright states, mediated by collective quantum jumps. Therefore, the system exhibits both metastability and bistabity, and will switch between the two states after reaching the stationary state.

In contrast, for the monostable II regime, the stationary state lies entirely within one of the two subspaces regardless of $N$ [$\Delta=4.4$; Figs.~\ref{fig:D3.4_vs_D4.4}(b) and (d)]. In this case, the system exhibits only metastability, and once it reaches the quantum dark state $\hat{\rho}_{+}$, it will remain there definitely. This suggests the possibility of constructing initial states whose relaxation times are significantly shorter than the inverse gap.  In a similar vein to the quantum Mpemba effect, where the accelerated relaxation is due to zero overlap between initial states and metastable modes~\cite{zhang2025observation,carollo2021exp}, the dynamic behavior of the system is strongly influenced by the choice of initial conditions. Specifically, because $D[\hat{\rho}_{+(-)},\hat{\rho}_\text{ss}]\approx 1(0)$, if we initialize the system with one of the eigenstates of $\hat{\rho}_{-}$ ($\hat{\rho}_{+}$), a speed-up (slowdown) of relaxation will occur.

\subsection{\label{sec:PDF}Steady-state occupation probabilities of metastable states}

\begin{figure}[b]
	\centering
	\includegraphics[width = 8.6 cm, keepaspectratio]{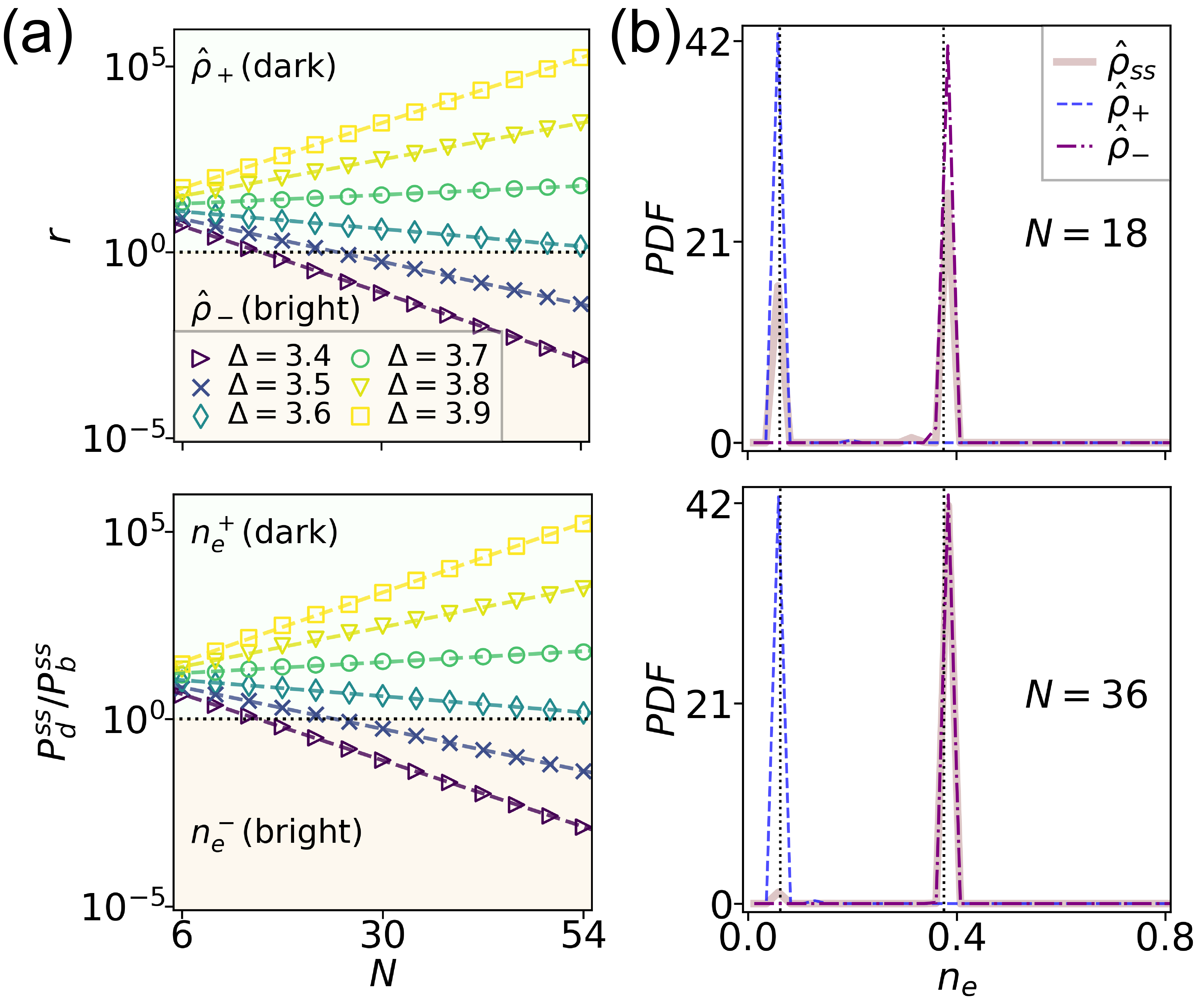}
	\caption{(a) Finite-size scaling of the ratio of the steady-state occupation probability of the dark state to that of the bright state obtained through (upper) projection of the stationary state to the metastable subspaces and (lower) rate-function approximation for various values of detuning $\Delta$.  Dashed lines are nonlinear fitting $b e^{a N}$. All curves focus on the bistable regime ($\Delta\approx3\sim4.2$). 
    (b) The probability distribution function (PDF) for the Rydberg population $n_e$ is associated with the density matrices $\hat{\rho}_\text{ss},\hat{\rho}_\pm$ for different system sizes in the bistable region with $\Delta=3.4$. Dotted vertical lines stand for the corresponding stable MF fixed points. }
	\label{fig:Prob}
\end{figure}

We have demonstrated that in the bistable regime, the stationary state emerges in the form of a statistical mixture of the quantum dark and bright states mediated through collective quantum jumps. It is possible to determine the relative probability of finding the system in one metastable state relative to the other in the bistable regime. We define the ratio $r=D[\hat{\rho}_+,\hat{\rho}_\text{ss}]/D[\hat{\rho}_{-},\hat{\rho}_\text{ss}]$ to characterize the relative probability, where $r\gg1$ ($r\ll1$) indicates the system predominantly in the dark (bright) state. As shown in Fig.~\ref{fig:Prob}(a), small systems prefer the dark state regardless of detuning as they tend to minimize the fluctuations~\cite{horsthemke1984noise,biancalani2012noise,biancalani2014noise,
jafarpour2015noise,jhawar2020noise}.
The ratio $r$ varies exponentially with the system size $N$, with an exponent that depends on the detuning. For small $\Delta$, the bright state is preferred over the dark one, whereas a gradual transition occurs toward dominance of the dark state when increasing $\Delta$. 

Meanwhile, the steady-state Rydberg excitation density $n_e^\text{ss}$ is given by
\begin{equation}
\label{eq:ne_t}
n_e^\text{ss}\approx n_e^+ P_\text{d}^\text{ss} + n_e^-P_\text{b}^\text{ss}
\end{equation}
where $P_\text{d(b)}^\text{ss}$ is the steady-state occupation probability of the dark (bright) state with $P^\text{ss}_\text{d}=(n_e^\text{ss}-n_e^{-})/(n_e^{+}-n_e^{-})$, $P^\text{ss}_\text{b}=1-P^\text{ss}_\text{d}$ and $n_e^{-(+)}$  is the Rydberg densities of the quantum bright (dark) state.
This reproduces the heuristic rate-function approximation, wherein the relaxation process is further simplified to transitions between the dark and bright states~\cite{savage1988single,wilson2016collective,brookes2021critical,fitzpatrick2017observation}. Within the bistable regime, the ratio of the steady-state occupation probabilities $P_\text{d}^{\text{ss}}/P_\text{b}^{\text{ss}}$ [lower panel of Fig.~\ref{fig:Prob}(a)] exhibit a similar exponential dependence on the system size as that of the probabilities of finding the system in the two subspaces [upper panel of Fig.~\ref{fig:Prob}(a)].

A finer structure of the steady state is accessible through the quantum-classical mapping. We begin by noting that the Markovian quantum master equation \eqref{eq:lindblad} can be unraveled in terms of quantum trajectories, which entails simulating a single quantum trajectory of the dissipative process \cite{belavkin1990stochastic,gardiner1992wave,dalibard1992wave,plenio1998the,carollo2019unravel}. The quantum-classical mapping involves considering each simulated quantum trajectory as the stochastic evolution of a pure state $\hat{\psi}_t\equiv\ket{\psi_t}\bra{\psi_t}$. After averaging over noise configurations, this yields the probability distributions $P_t(\hat{\psi})$ of finding the system in the pure state $\hat{\psi}$ at time $t$. Consequently, the density matrix is given by $\hat{\rho}_t=\int{d\hat{\psi} P_t(\hat{\psi})\hat{\psi}}$~\cite{carollo2019unravel}. 

Furthermore, since the expectation values of all observables can be evaluated as classical expectation values of the stochastic process of quantum jump trajectories, we construct a probability distribution function (PDF) from the density matrices $\hat{\rho}_{\alpha}$ ($\alpha=\text{ss}, \pm$).
This is achieved by employing direct diagonalization $\hat{\rho}_{\alpha}=\sum_l{P_l^{\alpha} \hat{\psi}_l^{\alpha}}$ with $\hat{\psi}_l^{\alpha}\equiv\ket{P_l^{\alpha}}\bra{P_l^{\alpha}}$ representing the $l$-th pure state of the density operator $\hat{\rho}_\alpha$, and the corresponding eigenvalue $P_l^{\alpha}$ is the probability of finding the system in that state. The associated discretized version of PDF for variable $x$ follows from data binning according to 
\begin{equation}
\text{PDF}(x;\alpha)\equiv \frac{\sum_l{P_l^{\alpha}\theta(x_{l}^\alpha-x+c)\theta(x+c-x_{l}^\alpha)}}{2c}
\end{equation}
where $x_{l}^\alpha = \bra{P_l^{\alpha}}\hat{x}\ket{P_l^{\alpha}}$ is the expectation value of observable $\hat x$ evaluated with the $l$-th pure state of the density operator $\hat\rho_\alpha$. Here we consider $x=n_e$, which represents the PDF for the Rydberg densities $n_e$.

The resultant PDF$(n_e; \alpha)$ is shown in Fig.~\ref{fig:Prob}(b). In the bistable region, for a given $\Delta$, there are two peaks of the steady-state distribution (red solid curves) located at the two stable MF fixed points with heights conditional upon the system size $N$, in alignment with the ratio $r$ of the occupation probabilities [see Fig.~\ref{fig:Prob}(a)]. In contrast to the bimodal distribution of the steady-state PDF, the PDF related to the two subspaces $\hat{\rho}_\pm$ (purple and blue curves) peaks only at one of the two MF fixed points regardless of $N$. This is consistent with the behavior of the expectation values $n_e^\alpha$ based on the density matrices $\hat{\rho}_{\text{ss}}$ and $\hat{\rho}_{\pm}$ [see Fig.~\ref{fig:D3.4_vs_D4.4}(a)]. 

As briefly alluded to in the last section, nonzero occupation probabilities of the two states have significant implications for the long-time dynamics. We refer to the quantum jump from the dark (bright) to bright (dark) states as upward (downward) and denote the corresponding switching rate as $\Gamma_d$ ($\Gamma_b$). The slowest relaxation process is described by
\begin{equation}
\partial_t\hat{\rho}(t)=\hat{\rho}_\text{ss}+A(\hat{\rho}_{+}-\hat{\rho}_{-}) e^{-\lambda_1 t}
\end{equation}
where coefficient $A$ depends on the initial condition. Each quantum trajectory represents a time record of pure states. Therefore, the nonzero occupation of both disjoint states $\hat{\rho}_\pm$ implies that, after each upward (downward) jump, coefficient $A$ is reset to $-D[\hat{\rho}_{+},\hat{\rho}_\text{ss}]$ ($D[\hat{\rho}_{-},\hat{\rho}_\text{ss}]$). This indicates that the quantum metastability is always at play. Accordingly, the switching rates $T_b^{-1}$ and $T_d^{-1}$ are proportional to $-\lambda_1 P_\text{d}^\text{ss}$ and $-\lambda_1 P_\text{b}^\text{ss}$, respectively. The steady state in bistable systems is essentially the dynamic mixing of the dark and bright states mediated by exponentially rare collective jumps. Consequently, the combined effects of metastability and bistability bring about dynamic hysteresis in large systems~\cite{casteels2016power,rodriguez2017probing,chen2023quantum,carr2013Nonequilibrium,ding2020phase}. 

In the next section, we will explicate how the exponential dependence of the relative occupation probabilities on $N$ arises from an Arrhenius-type exponential scaling with the effective energy barrier for escaping from the two metastable states.

\section{\label{sec:switching}Stochastic quantum switching}

\subsection{\label{sec:instanton}The instanton approach}

For Lindbladian dynamics, the quantum noises stem from the corresponding quantum Langevin equations \cite{gardiner1992wave,Gardiner1985input}, where on top of the deterministic dynamics generated by the effective non-Hermitian Hamiltonian, there are additional stochastic dynamics induced by the quantum noise operators. To bypass the difficulty in dealing with operators, we consider the semiclassical limit, where the deterministic parts are the same as Eqs.\eqref{eq:dmxdt}-\eqref{eq:dmzdt}, whilst the noise operators are replaced with their respective classical counterparts. The resultant Langevin equations can be cast into a dynamic path integral via a Martin-Siggia-Rose construction \cite{Martin1973,Janssen1976,Tauber2014critical} with the partition function
\begin{equation}
\label{eq:MSR_partition}
\mathcal{Z}=\int{\mathcal{D}\left[m_x,m_y,m_z,\tilde m_z,\tilde m_y, \tilde m_z\right] e^{-\mathcal{S}}}
\end{equation}
with the nonequilibrium action $\mathcal{S}$ (the sum over repeated indices is implied hereinafter),
\begin{equation}
\label{eq:MSR_action}
\mathcal{S} =\int{dt \left\{ \tilde{m}_\alpha \left(\partial_t m_\alpha - F_\alpha - \frac{M_{\alpha\beta}}{2}\tilde{m}_\beta\right)\right\}}
\end{equation}
where $\alpha,\beta\in\Set{x,y,z}$, the generalized force $F_\alpha$ is given by Eq.~(\ref{eq:EOM_MF}), and the covariance $M_{\alpha\beta}$ for the white, zero-mean Langevin noise $\eta_{\alpha,t}$ is fixed via the generalized Einstein relation~\cite{pan2020non,marcuzzi2016absorbing,buchhold2017nonequilibrium,helmrich2020signatures}, i.e., $\expval{\eta_{\alpha,t}\eta_{\beta,t'}}=\frac{\delta(t-t')}{2N}\expval{\partial_t(\hat{S}^\alpha_t\hat{S}^\beta_t)-\partial_t(\hat{S}^\alpha_t)\hat{S}^\beta_t-\hat{S}^\alpha_t\partial_t(\hat{S}^\beta_t) + \text{h.c.}}$, with h.c. representing the Hermitian conjugate (see Sec. II of SM for details). 
By introducing the vectorial notation 
$\va{\eta}_t=
\begin{pmatrix}
\eta_{x,t},&\eta_{y,t},&\eta_{z,t}
\end{pmatrix}^T$, the covariance matrix $\va{M}_t\delta(t-t')=\expval{\va{\eta}_t\va{\eta}^T_{t'}}$ is explicitly written as (time-dependence made implicit)
\begin{equation}
\va{M}=\frac{1}{N}
\begin{pmatrix}
1&0&m_x\\
0&1&m_y\\
m_x&m_y&2(m_z+1)
\end{pmatrix}
\end{equation}
By including the response fields $\tilde{m}_\alpha$, the phase space dimension increases from $3$ to $6$. We now introduce the momentum $q_\alpha=\tilde{m}_\alpha/N$ conjugate to $m_\alpha$ and assign a vector 
$\va{x}=
\begin{pmatrix}
m_x,&m_y,&m_z,&q_x,&q_y,&q_z
\end{pmatrix}^T$ 
to each state in phase space. 

The optimal transition trajectory is obtained through the saddle-path approximation to the action~\eqref{eq:MSR_action} with starting and ending points fixed~\cite{landau1976mechanics,coleman1988quantum,graham1986noneq,park2021stochastic,bressloff2014path,chernykh2001large,touchette2009large,kamenev2023field,elgart2004rare}.
This leads to the Hamilton-Jacobi equations for coordinates $m_\alpha$ and momenta $q_\alpha$ 
\begin{subequations}\label{eq:Hamilton_Jacobi}
\begin{align}
&\partial_t m_\alpha = \partial_{q_\alpha}H_\text{cl},\\
&\partial_t q_\alpha = - \partial_{m_\alpha}H_\text{cl}
\end{align}
\end{subequations}
where the (classical) Hamiltonian $H_\text{cl}$ is given by
\begin{equation}
\label{eq:Hamiltonian}
H_\text{cl}=F_\alpha q_\alpha + \frac{1}{2}\bar{M}_{\alpha\beta} q_\alpha q_\beta
\end{equation}
with the rescaled noise covariance $\bar{M}_{\alpha\beta}=NM_{\alpha\beta}$. 

After inserting Eqs.~\eqref{eq:Hamilton_Jacobi} and~\eqref{eq:Hamiltonian} to the action~\eqref{eq:MSR_action}, the action associated with transition from state $\va{x}_a$ at time $t=0$ to state $\va{x}_b$ after a finite time $T\geq0$ reads
\begin{equation}
\label{eq:action_increment}
\mathcal{S}_{T} = N \int_0^T{dt \left( q_\nu \partial_t m_\nu - H_\text{cl} \right)}=\frac{N}{2}\int_0^T{dt \left( q_\nu M_{\nu\alpha}q_\alpha\right)}
\end{equation}
which vanishes for deterministic dynamics ($q_\alpha=0$) and is positive when fluctuations are needed ($q_\alpha\neq0$) as is the case for switching between metastable states.
\begin{figure}[t]
	\centering
	\includegraphics[width = 8.6 cm, keepaspectratio]{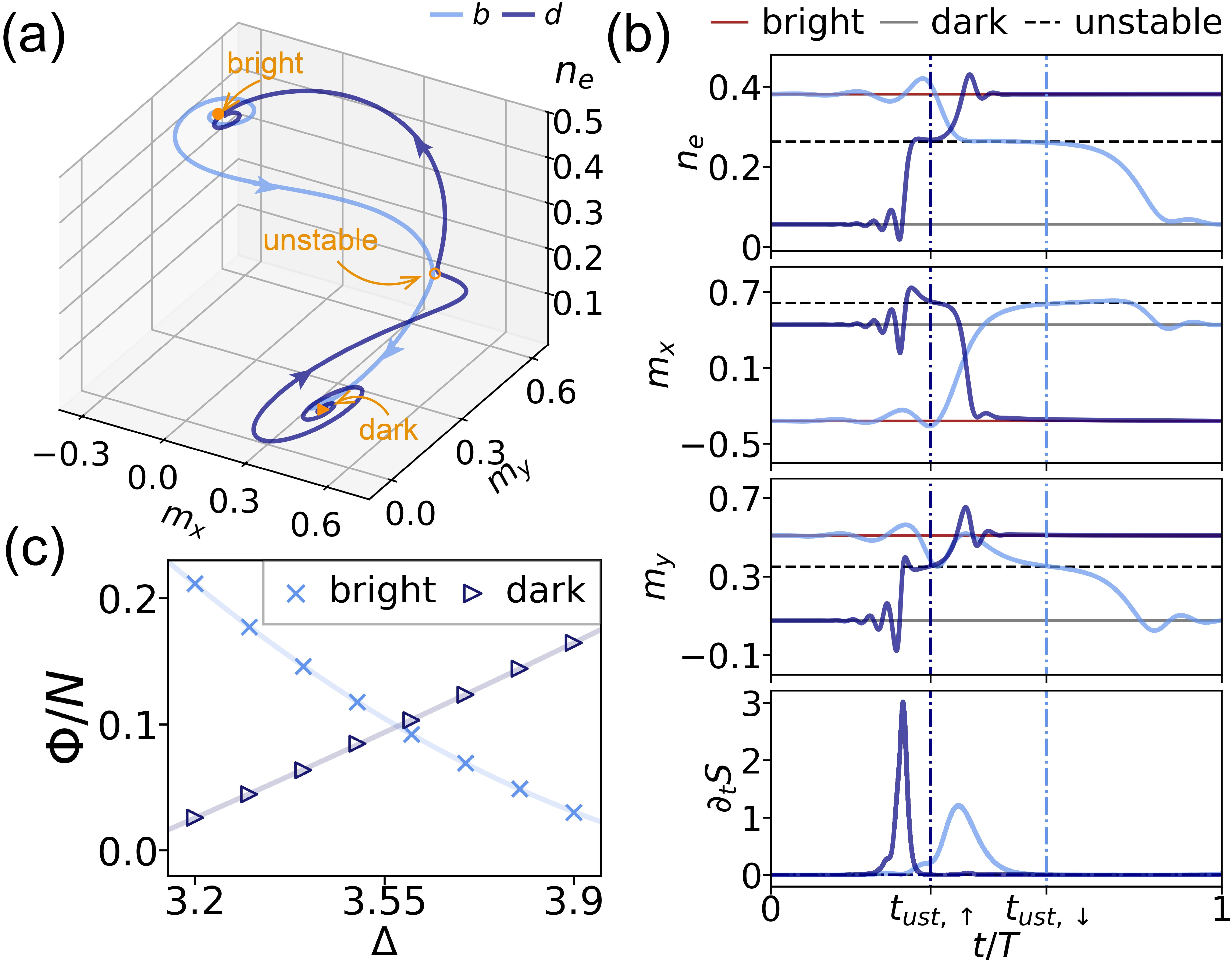}
	\caption{Optimal switching paths (instantons) obtained within the Martin-Siggia-Rose dynamic path integral formalism after semiclassical approximation. 
    (a) The optimal paths for switching from the dark (bright) state to the bright (dark) state for $\Delta=3.5$. The dot, triangle, and circle represent the dark, bright, and unstable states, respectively. The arrows along the curves indicate the direction of time.
    (b) The spin fields $n_e, m_x, m_y$, and the action increment rate along the optimal paths in (a) as a function of rescaled time $t/T$, where $T$ is the total evolution time. The dashed (solid) horizontal lines denote the unstable (stable) MF fixed point(s), while the vertical lines indicate the times at which the systems reach the unstable state.
    (c) The effective energy barrier (quasipotential) $\Phi$, defined as the increment of action along the optimal path divided by the system size $N$ as a function of detuning $\Delta$ for dark and bright states.}
	\label{fig:Instanton}
\end{figure}
Because the Hamiltonian~\eqref{eq:Hamiltonian} is an integral of motion, the action~\eqref{eq:action_increment} can be parameterized with ``energy'' $E$
\begin{equation}
\label{eq:action_increment_energy}
\mathcal{S}_{T,E} =N \left( \int_{a}^{b}{q_\nu d m_\nu}-ET\right)
\end{equation}

We also obtain the effective energy barrier (quasipotential) $\Phi(\va{x}_a\to\va{x}_b)$ from state $\va{x}_a$ to $\va{x}_b$, which is defined as the accumulated action through the optimal paths (the so-called instantons)
\cite{coleman1988quantum,Dykman1994LargeFA,kamenev2023field,elgart2004rare,graham1986noneq,park2021stochastic,bressloff2014path,chernykh2001large,touchette2009large,zakine2023minimum},
\begin{equation}
\label{eq:erg_barrier}
\Phi(\va{x}_a\to\va{x}_b) \equiv \inf_{T>0}{\inf{\mathcal{S}_{T,E}}}
\end{equation}
where the minimization is taken over all paths ${\va{x}(t)}$ with the boundary conditions $\va{x}(0)=\va{x}_a$, $\va{x}(T)=\va{x}_b$ ($t\in[0,T]$). It follows from Eq.~\eqref{eq:MSR_partition} that the asymptotic transition rate $\Gamma_{\va{x}_a\to\va{x}_b}$ from state $\va{x}_a$ to $\va{x}_b$ takes the LD form
\begin{equation}
\Gamma_{\va{x}_a\to\va{x}_b} \approx e^{-\Phi(\va{x}_a\to\va{x}_b)} 
\end{equation}
 
As indicated by Eq.~\eqref{eq:action_increment_energy}, the action scales linearly with the system size, thereby indicating an exponential scaling as discussed above. As a result, the rate $\Gamma_{\va{x}_a\to\va{x}_b}$ asymptotically approaches $0$ ($1$) as $N$ approaches infinity (zero). Additionally, the quasipotential $\Phi$ is time-independent, implying that the optimal paths are also time-invariant. Since the time $T$ in Eq.~\eqref{eq:action_increment_energy} is not fixed, the optimal path can be rendered stationary only when $E=0$~\cite{coleman1988quantum,kamenev2023field,elgart2004rare}. 
The three fixed points $\va{x}_b$, $\va{x}_d$, and $\va{x}_u$ which nullify Eq.~\eqref{eq:Hamilton_Jacobi} are located on the zero-energy manifold and correspond to the MF bright, dark, and unstable states, respectively. Consequently, a trajectory that connects any two of them and evolves with time according to Eqs.~\eqref{eq:Hamilton_Jacobi} is the optimal path. 

We utilize the minimum-action method~\cite{zakine2023minimum} to find the optimal path for switching. The resulting trajectories of upward and downward jumps, which correspond to the optimal paths for escaping from the dark and bright states, are displayed in Fig.~\ref{fig:Instanton}(a). The time series of the spin fields and the action increment rate $\partial_t\mathcal{S}$ associated with these paths are shown in Fig.~\ref{fig:Instanton}(b). As indicated by Eq.~\eqref{eq:action_increment}, escaping from the respective initial attractive basins requires nonzero fluctuations, thereby resulting in action increment for both upward and downward jumps. Upon reaching the unstable fixed point (highlighted with the vertical dashed lines), the system relaxes towards another attractive basin along the deterministic paths, which are devoid of fluctuations and action increments~\cite{Dykman1994LargeFA}. 

The effective energy barrier normalized by the system size calculated according to Eqs.~\eqref{eq:action_increment_energy} and~\eqref{eq:erg_barrier} is plotted in Fig.~\ref{fig:Instanton}(c). It is evident that the energy barrier for the bright (dark) state decreases (increases) with increasing detuning $\Delta$. This trend results in an increasing (decreasing) escaping rate for the bright (dark) state. Such behavior is consistent with the increase in the ratio of the steady-state occupation probability of the dark state to that of the bright state [see Figs.~\ref{fig:Prob}(a)].

\subsection{\label{sec:QJMC}Quantum-jump simulations}

\begin{figure}[t]
	\centering
	\includegraphics[width = 8.3 cm, keepaspectratio]{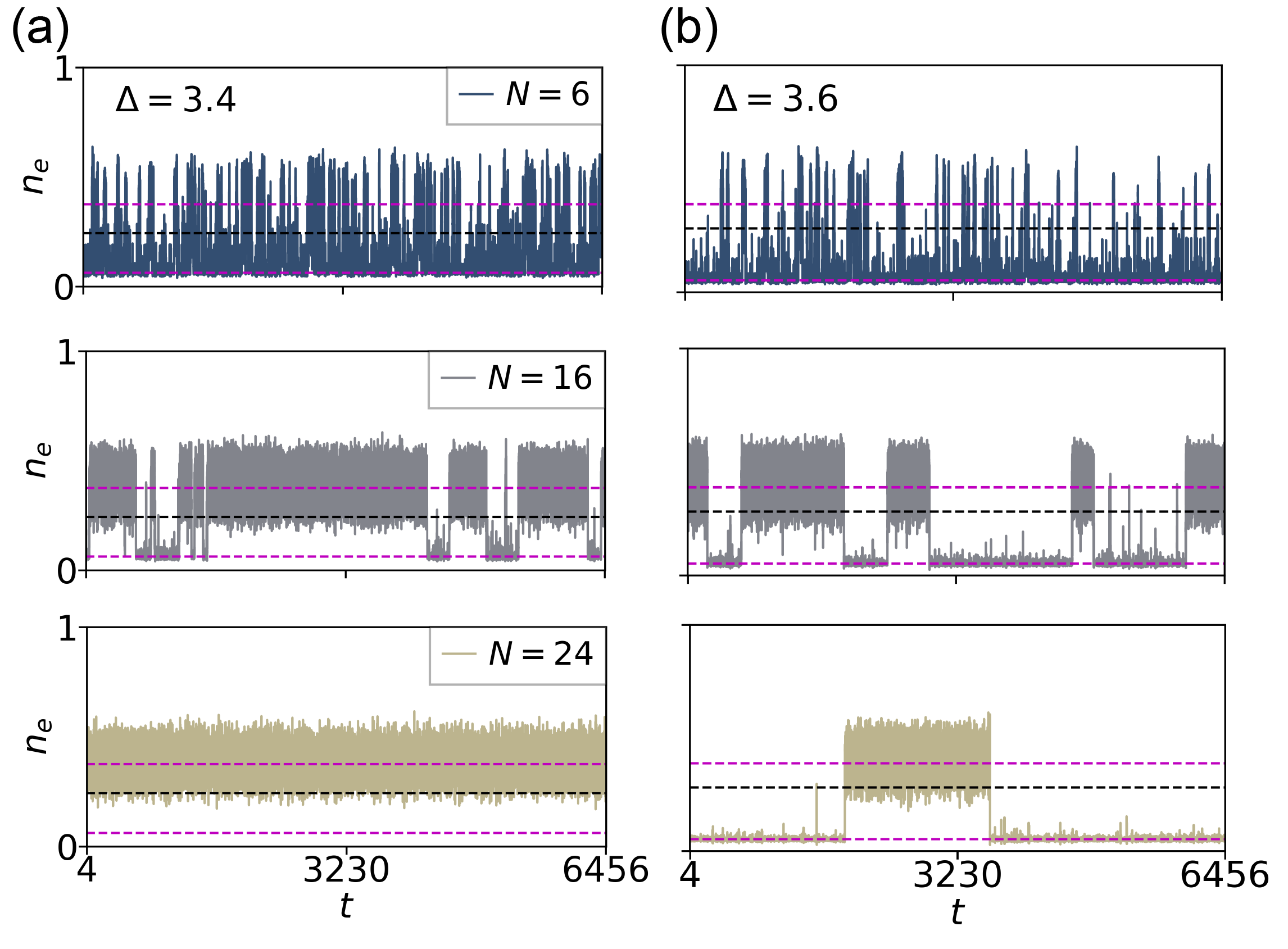}
	\caption{Average excitation population of simulated quantum trajectory for (a) $\Delta=3.4$ and (b) $\Delta=3.6$ of varying number of atoms $N$ over time. The gray (magenta) dashed lines denote the unstable (stable) MF fixed point(s).}
	\label{fig:Trj}
\end{figure}

The classical path integral predicts an exponential dependence of waiting times for collective quantum jumps on the system size. This implies that signatures of first-order transitions can be probed at the level of quantum trajectories. In addition, since errors in the form of noises and quasipotentials can lead to exponentially large discrepancies in the waiting times, we resort to simulating quantum trajectories to improve accuracy. By keeping track of the upward and downward jumps from these simulated trajectories, we extract the corresponding waiting times. In the simulation, given the time step $dt$, at each step the wave function $\ket{\psi_t}$ either collapses $\ket{\psi_t}\to{\hat{L}\ket{\psi_t}}/{\sqrt{\bra{\psi_t}\hat{L}^{\dagger}\hat{L}\ket{\psi_t}}}$ with a probability of $P_t=dt\bra{\psi_t}\hat{L}^{\dagger}\hat{L}\ket{\psi_t}$, or evolves under the action of the effective non-Hamiltonian $\ket{\psi_t}\to {e^{-i\hat{H}_\text{eff}dt}\ket{\psi_t}}/{\sqrt{1-P_t}}$ with a probability $1-P_t$. The time-dependent average Rydberg population is computed according to $n_e(t)=\bra{\psi_t}\hat{S}^z\ket{\psi_t}/N+0.5$~\cite{gardiner1992wave,dalibard1992wave,plenio1998the}. 

Exemplary quantum trajectories are displayed in Fig. \ref{fig:Trj}, where it is evident that for small $N$, the system is mainly trapped in a dark state and occasionally experiences significant quantum fluctuations from time to time (upper panels). In contrast, the wave function jumps back and forth between the two states (central panels) in larger systems. Relative to the time spent trapped around either of the two states, the jump appears almost instantaneously, and the waiting times between successive jumps increase with $N$, indicating elongated lifetimes of the two states. This is consistent with the smaller gap of the Liouvillian spectrum. Additionally, in larger systems, as $\Delta$ increases from $3.4$ to $3.6$, the waiting times for upward jumps surpass those for downward jumps [central and lower panels in Fig.~\ref{fig:Trj}(a) and (b)]. This suggests that the dark (bright) state is more stable than the other for large (small) detuning in the bistable region, consistent with
the spectral decomposition results shown in Fig.~\ref{fig:Prob}. 

\begin{figure}[t]
	\centering
	\includegraphics[width = 8.3 cm, keepaspectratio]{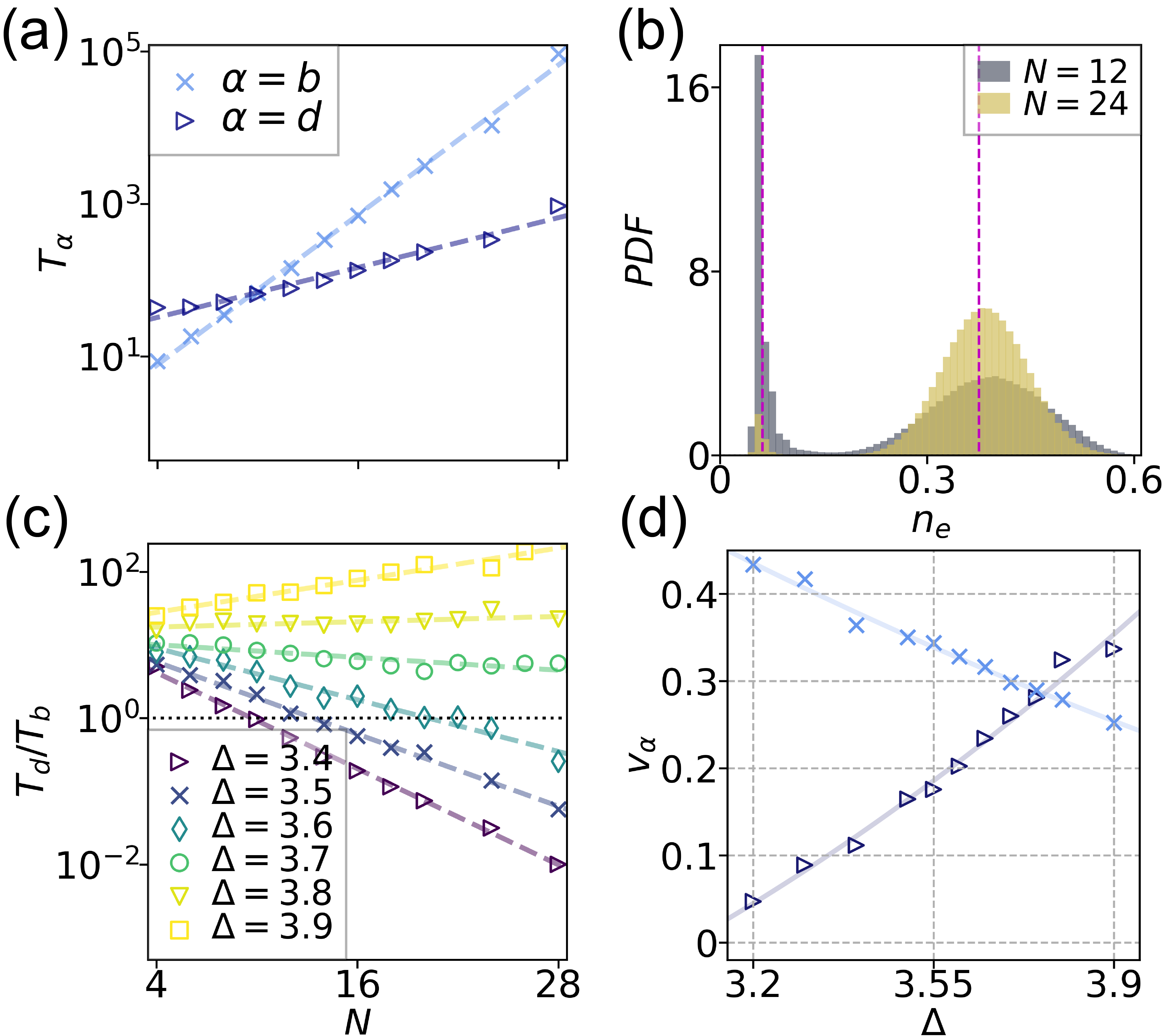}
	\caption{Mean waiting times from quantum-jump simulations.
	(a) Mean waiting times for the dark  $T_d$ and bright $T_d$ states as a function of number of atoms $N$ for $\Delta = 3.4$.
	(b) Probability distributions of the average Rydberg population of simulated trajectories for $\Delta=3.4$ and $N=12,24$. 
    (c) Ratio of the mean waiting times for upward to downward jumps for varying detuning and system size.
    (d) The effective energy barrier $v_{d(b)}$ for the dark (bright) states extracted from fitting the waiting times with ($\alpha=b,d$) $T_\alpha=b_\alpha e^{v_\alpha N}$, where the exponent $v_\alpha$ corresponds to the slope of curves in panel (a).}
	\label{fig:Jump}
\end{figure}
\begin{figure}[t]
	\centering
	\includegraphics[width = 6.5 cm, keepaspectratio]{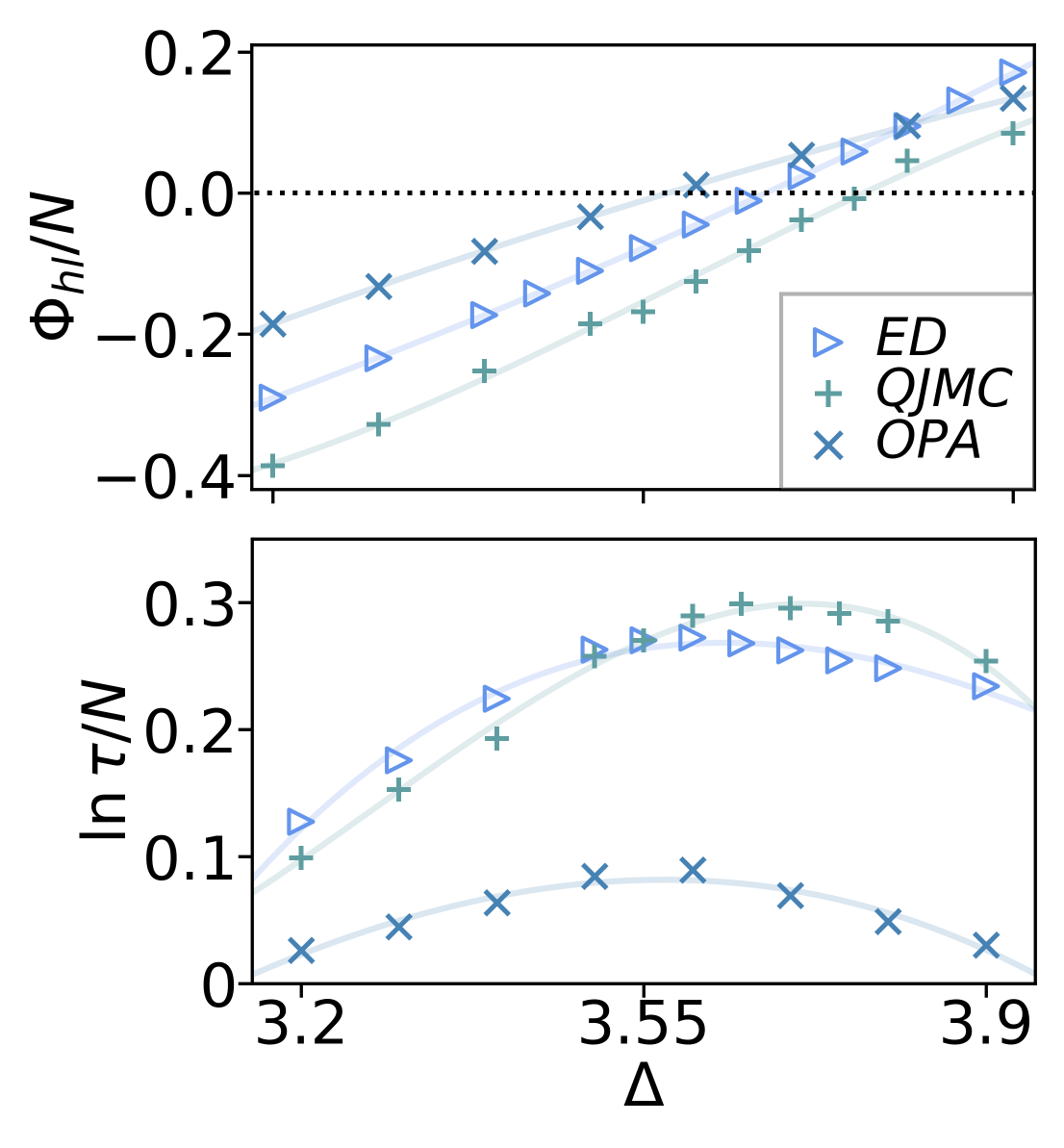}
	\caption{(a) The difference in the normalized effective energy barrier $\Phi_{db}$ between the dark and bright states and (b) the relaxation time $\tau$ within the bistable regime estimated through the exact diagonalization (ED) of the Liouvillian, the quantum-jump Monte-Carlo (QJMC) simulations, and the optimal path approximation (OPA) as a function of detuning $\Delta$.}
	\label{fig:barrier_compare}
\end{figure}

A comparison of the mean switching times $T_d$ and $T_b$ across different numbers of atoms $N$ with fixed detuning $\Delta = 3.4$ is given in Fig.~\ref{fig:Jump}(a). Both $T_d$ and $T_b$ increase exponentially with $N$, albeit with different exponents. Consequently, although the dark state is generally preferred for small $N$ due to reduced fluctuations, the waiting time (lifetime) of the bright state increases with $N$ more rapidly than that of the dark state, indicating that the bright state becomes more stable in larger systems. In accord, the PDF of the average Rydberg population $n_e(t)$ is bimodal and peaks at the two metastable states [Fig.~\ref{fig:Jump}(b)]. As $N$ increases, so does the height of the peak that corresponds to the bright state, consistent with the longer lifetime of the bright state compared to the dark state. 

As evinced in Fig.~\ref{fig:Jump}(c), parallel to the occupation probabilities [see Fig.~\ref{fig:Prob}(a)], the ratio of the two switching rates $T_d/T_b$ also varies exponentially with the system size, with a deviation associated with the uncertainties of determining the bright and dark states from fluctuating data. The relation between the steady-state occupation probabilities and the waiting times of the two quantum states reflects the essence of their dynamical coexistence, providing the physical basis for the emergence of a unique steady state from two metastable states.
Through log-linear fitting of the waiting times  $T_\alpha \propto e^{v_\alpha N}$ ($\alpha=b,d$), we can estimate the effective energy barrier $v_\alpha$ for each state. It is seen from Fig.~\ref{fig:Jump}(d) that the energy barrier for the dark (bright) state increases (decreases) with the detuning, in agreement with the results based upon the optimal paths [see Fig.~\ref{fig:Instanton}(c)]. 

Fig.~\ref{fig:barrier_compare}(a) shows the difference in the normalized effective energy barrier between the dark and bright states, calculated from the occupation probabilities of two respective subspaces $\hat{\rho}_\pm$, mean waiting times, and the action increments along the optimal path.
They demonstrate the connection between the steady-state occupation probabilities and the mean lifetimes of the metastable states, which are captured in terms of the varied effective quasipotential from one state to the other. The relaxation time dictated by the switching rates is estimated through~\cite{wilson2016collective} $\tau\approx(T_b^{-1}+T_d^{-1})^{-1}\propto e^{\max{\{v_b,v_d\}} N}$, which also shows exponential size scaling. As plotted in Fig.~\ref{fig:barrier_compare}(b), the exponent $\ln{\tau}/N$ reaches a peak around $\Phi_{bd}\approx 0$ , where both states exist almost on an equal footing. It decreases as the difference between the two states increases and vanishes outside the bistable regime.

\section{\label{sec:conclusion}Conclusion and discussion}
In this work, we systematically investigate the connection between quantum bistability and collective quantum jumps in a system exhibiting a first-order dissipative phase transition in the thermodynamic limit. We demonstrate that, in finite systems far from the thermodynamic limit, discontinuous phase transitions emerge through quantum metastability with distinct features in both statistics of quantum fluctuations and the low-lying eigenmodes of the Liouvillian. 

We find that quantum bistability has its origin not only in the closing of the spectral gap but also in the nonzero steady-state partition of disjoint metastable subspaces. The former is a signature of quantum metastability at the spectrum level, whereas the latter indicates quantum metastability at the trajectory level, which is characterized by stochastic switching between distinct metastable states. This demonstrates that bistable systems are essentially metastable systems where the unique stationary state is sustained by the rare collective jumps between the two metastable states. In contrast, systems displaying metastability without bistability merely display a slowed relaxation toward the stationary state when metastable modes are initially excited.

Our results generalize the Arrhenius scaling of the decay rates of homogeneous metastable states to quantum systems, establishing a connection between classical stochastic dynamics and quantum bistability. Crucially, the sheer quantum origin distinguishes the collective jumps from their classical counterparts. Occurring at zero absolute temperature, the stochastic switching is purely driven by quantum fluctuations. Like their equilibrium counterparts, the two metastable states differ in their robustness to fluctuations, which is manifested in their steady-state occupation probabilities and switching rates. Through a semiclassical instanton approach and quantum-jump simulations, we reveal that the mean times between successive jumps diverge exponentially with system size, with different exponents reflecting the effective potential barriers for each state. This exponential size scaling is absent in spatially extended systems, where escape from the false to the true vacua is limited by critical nuclei (bubbles) that are independent of system sizes~\cite{cates2023classical,lagnese2024detecting}. 

The relation between switching kinetics and steady-state occupation probabilities allows the identification of the critical parameter values for a first-order phase transition, where the switching between the two states occurs at the same timescales and the two states coexist on equal terms. This approach not only circumvents diagonalizing the Liouvillian but, more importantly, provides a criterion to distinguish bistability from metastability in finite systems. The important and often neglected distinction between metastability at the spectral and trajectory levels derives from two complimentary approaches to open quantum systems: the deterministic equation for the density matrix and the stochastic equation for the quantum state. While the evolution of density matrices is completely determined by initial conditions, stochastic quantum trajectories quickly converge into a metastable state. This leads to a hallmark of quantum bistability: the relaxation of quantum trajectories can slow down significantly compared to their corresponding density matrices.

Our analysis is not restricted to microscopic details and applies generally to homogeneous multistable systems. Given the enduring interest in the switching phenomenon and its potential applications, it is important to consider and exploit the exponential size scaling in quantum multistable systems. For example, by increasing (decreasing) the system size, signals with lower (higher) frequencies can be detected via stochastic resonance~\cite{gammaitoni1998sr}.

\section*{Acknowledgment}
We acknowledge financial support by the National Natural Science Foundation of China under Grants No. 12274131, No. 12347102 and No. 12174184, the Natural Science Foundation of Jiangsu Province under Grant No. BK20233001, and the Innovation Program for Quantum Science and Technology under Grant No. 2024ZD0300101. W.L. acknowledges financial support from the EPSRC (Grant No. EP/W015641/1). We also acknowledge computational resources provided by the High Performance Computing Center of Nanjing University.

\normalem

%

\end{document}